\def\db#1{\mathbb #1}
\def\cal#1{{\mathcal #1}}
\def\esm#1{{\db E\left[\,#1\, \right]}}
\def\cnd{{\,|\,}}
\def\pr#1{{\db P\left\{\,#1\,\right\}}}

\def\mathQED{{_\blacksquare}}
\def\eps{\epsilon}
\def\one{{\bf 1}}

\def\ux{{{\underline x}}}
\def\pn{\par\noindent}
\def\psn{\par\smallskip\noindent}
\def\pmn{\par\medskip\noindent}

\def\Lam{\Lambda}
\def\om{{\omega}}

\def\cX{{\cal X}}

\def\cF{{\cal F}}

\def\cH{{\cal H}}
\def\cS{{\cal S}}

\def\pt{\partial}

\def\pn{\par\noindent}
\def\prop#1{{{\pn\bf#1. }}}

\def\proof{{\pmn\prop{Proof}}}

\def\dist{{{\rm dist}}}

\def\card{{{\rm card}}}
\def\supp{{{\rm supp}}}
\def\myset#1{{\left\{\, #1 \, \right\}}}

\def\Z{{\db Z}}
\def\R{{\db R}}

\def\truc#1#2#3{\smash{\mathop{\,\, #1 \,\, }\limits^{#2}_{#3}}}

\documentclass[notitlepage]{article}
\usepackage[a4paper]{geometry}
\usepackage[intlimits]{amsmath}
\usepackage{amssymb}
\usepackage{amsfonts}
\usepackage{theorem}
\newtheorem{Thm}{Theorem}
\newtheorem{Def}{Definition}
\newtheorem{Lem}{Lemma}
\newtheorem{Rem}{Remark}

\begin{document}
\title{UNIVERSITE DE REIMS CHAMPAGNE-ARDENNE
   \\ Laboratoire de Math\'{e}matiques \\ UMR 6056
\vskip 5 truecm
 A SIMPLE EXTENSION OF STOLLMANN'S LEMMA \\ TO CORRELATED POTENTIALS }
\author{ Victor Chulaevsky$^1$ }

\date{}
\maketitle


\vfill
\pn
------------------------------------------------------------------------------------------------------------------------
\noindent
\begin{center} $^1$ {\bf Permanent address:}
 Victor Tchoulaevski \\
Universit\'{e} de Reims, Laboratoire de Math\'{e}matiques\\
 Moulin de la Housse - B.P. 1039 - 51687 Reims cedex 2 \\ France
\pn
E-Mail: victor.tchoulaevski@univ-reims.fr
\end{center}
\pn
\vskip 1 truecm
\newpage

\begin{center}
   {\bf A SIMPLE EXTENSION OF STOLLMANN'S LEMMA \\ TO CORRELATED POTENTIALS }

\vskip 1 truecm
 {\bf Victor Chulaevsky}
\medskip
 \\ D\'{e}partement de Math\'{e}matiques, UMR CNRS 6056 \\
 Universit\'{e} de Reims, Moulin de la Housse - B.P. 1039 \\ 51687 Reims cedex 2,  France
 \\ E-mail: victor.tchoulaevski@univ-reims.fr
\end{center}
\bigskip

\abstract
  We propose a fairly simple and natural extension of Stollmann's lemma to
correlated random variables, described earlier in \cite{Ch}. This extension allows
(just as the original Stollmann's lemma does) to obtain  Wegner-type estimates even in
some problems of spectral analysis of random operators where the Wegner's lemma is
inapplicable (e.g. for multi-particle Hamiltonians).

 To the best of author's knowledge, such an extension seems
to be original. However, the author will appreciate any reference to articles or
preprints where  similar results are proved.

\pmn

\section{Introduction}

   The regularity problem for the  limiting distribution of eigen-values of infinite dimensional
self-adjoint operators appears in many problems of mathematical physics. Specifically,
consider a lattice Schr\"{o}dinger operator (LSO, for short) $H:\ell^2(\Z^d)\to \ell^2(\Z^d)$
given by
$$
(H\psi)(x) = \sum_{y:\, |y-x|=1} \,\, \psi(y) + V(x)\psi(x); \; x,y\in\Z^d.
$$
For each finite subset $\Lam\subset \Z^d$, let $E_j^\Lam, j=1,\dots, |\Lam|$, be
eigen-values of
$H$ with Dirichlet b.c. in $\Lam$. Consider the family of finite sets
$\Lam_L = [-L,L]^d \cap \Z^d$ and define the following quantity (which does not
necessarily exist for an arbitrary LSO):
$$
k(E)
= \lim_{L\to\infty} \frac{1}{(2L+1)^d} \card \left\{ j:\,E_j^{\Lam_L}\leq E \right\}.
$$
If the above limit exists, $k(E)$ is called the limiting distribution function (LDF) of
e.v. of $H$. One can easily construct various examples of the function $V:\Z^d\to \R$
(called potential of the operator $H$) for which the LDF does not exist. One can prove
the existence of LDF for periodic potentials $V$, but even in this, relatively simple
situation existence of $k(E)$ is not a trivial fact.

   However, one can prove existence of $k(E)$ in a large class of {\it ergodic random}
potentials. Namely, consider an ergodic dynamical system
$(\Omega,\cF,\db P, \{T^x, \, x\in\Z^d\})$ with discrete time $\Z^d$ and a mesurable
function (sometimes called a hull) $v:\Omega\to\R$. Then we can introduce a family  of
sample potentials
$$
V(x,\omega) = v(T^x\omega), \; x\in\Z^d,
$$
labeled by $\omega\in\Omega$. Under the assumption of ergodicity of $\{T^x\}$, the
quantity
$$
k(E,\omega)
= \lim_{L\to\infty} \frac{1}{(2L+1)^d} \card \left\{ j:\,E_j^{\Lam_L}(\omega)\leq E \right\}
$$
is well-defined $\db P$-a.s. Moreover, $k(E,\omega)$ is $\db P$-a.s. independent of
$\omega$, so its value taken for a.e. $\omega$ is natural to take as $k(E)$. In such a
context, $k(E)$ is usually called {\it integrated density of states} (IDS, for short). It
admits an equivalent definition:
$$
k(E) = \esm{ (f,\Pi_(-\infty,E](H(\omega)f)},
$$
where $f\in\ell^2(\Z^d)$ is any vector of unit norm, and $\Pi_(-\infty,E](H(\omega)$ is
the spectral projection of $H(\omega)$ on $(-\infty,E]$. The reader can find a detailed
discussion of the existence problem of IDS in excellent monographs by Carmona and Lacroix
\cite{CL} and by Pastur and Figotin \cite{PF1}.

     It is not difficult to see that $k(E)$ can be considered as the distribution
function of a normalized measure, i.e. probability measure, on $\R$. If this measure
$dK(E)$, called measure of states, is absolutely continuous with respect to Lebesgue
measure $dE$, its density (or Radon--Nikodim derivative) $dK(E)/dE$ is called the density
of states (DoS). In physical literature, it is customary to neglect the problem of
existence of such density, for if $dK(E)/dE$ is not a function, then "it is simply a
generalized function". However, the real problem is not terminological. The actual,
explicit  estimates of the probabilities of the form
$$
\pr{\exists \, \text{ eigen-value } E^{\Lam_L}_j\in(a,a+\eps)}
$$
for LSO $H_{\Lam_L}$ in a finite cube $\Lam_L$ of size $L$, for small $\eps$, often
depend essentially upon the existence and the regularity properties of the DoS
$dk(E)/dE$.

   Apparently, the first fairly general result relative to existence and boundedness of
the DoS is due to Wegner \cite{W}. Traditionally referred to as Wegner's {\it lemma}, it
certainly deserves to be called {\it theorem}.

\begin{Thm}[Wegner] Assume that $\{V(x,\omega), x\in\Z^d\}$ are i.i.d. r.v. with bounded
density $p_V(u)$ of their common probability distribution: $\|p_V\|_\infty = C < \infty$.
Then the DoS $dk(E)/dE$ exists and is bounded by the same constant $C$.
\end{Thm}

   The proof can be found, for example,  in the  monograph \cite{CL}.

   This estimate and some of its generalizations have been used in the multi-scale
analysis (MSA) developed in the works by Fr\"{o}hlich and Spencer \cite{FS}, Fr\"{o}hlich,
Spencer, Martinelli ans Scoppola \cite{FMSS}, von Dreifus and Klein \cite{DK1},
\cite{DK2}, Aizenman and Molchanov \cite{AM}, and in a number of more recent works
where the so-called Anderson Localization phenomenon has been observed. Namely, it has
been proven that all e.f. of random lattice Schr\"{o}dinger operators decay exponentially at
infinity with probability one (for $\db P$-a.e. sample of random potential $V(\omega)$).
Von Dreifus and Klein \cite{DK2} proved an analog of Wegner estimate and used it in their
proof of localization for Gaussian and some other correlated (but non-deterministic)
potentials. The author of these lines recently proved, in a joint work with Yu. Suhov
\cite{ChS}, an analog of Wegner estimate for a system of two or more interacting quantum
particles on the lattice under the assumption of analyticity of the probability density
$p_V(u)$, using a rigorous path integral formula by Molchanov (see a detailed discussion
of this formula in the monograph \cite{CL}). In order to relax the analyticity assumption
in a multi-particle context, V.C. and Yu. Suhov later used (\cite{ChS2}) a more general
and flexible result guaranteeing existence and boundedness of the DoS: the Stollmann's
lemma, which we discuss below.

   In the present work, we propose a fairly simple and natural extension of Stollmann's
lemma to correlated, but still non-deterministic random fields generating random
potentials. {\it To the best of author's knowledge, such an extension seems to be
original, although very simple. However, the author will appreciate any reference to
published papers or preprints where the same or similar result was mentioned and proved.}
Our main motivation here is to lay out a way to interesting applications to localization
problems for multi-particle systems.

\section{Stollmann's lemma for product measures}

   Recall the Stollmann's lemma and its proof for independent r.v. Let
$m\geq 1$ be a positive integer, and $J$ an abstract finite set with $|J| (=\card J) = m$.
Consider the Euclidean space $\db R^J \cong \R^m$ with standard basis
$(e_1, \dots, e_m)$, and its positive quadrant
$$
\db R^J_+ = \myset{ q\in\db R^J:\, q_j\geq 0, \,\, j=1, 2, \dots, m }.
$$
For any measure $\mu$ on $\db R$, we will denote by $\mu^m$ the product measure $\mu
\times \dots \times \mu$ on $\db R^J$. Furthermore, for any probability measure $\mu$ and
for any $\eps>0$, define the following quantity:
$$
s(\mu, \eps) = \sup_{a\in\db R} \,\, \int_{a}^{a+\eps} d\mu(t)
$$
and assume that $s(\mu,\eps)$ is finite. Furthermore, let $\mu^{m-1}$ be the marginal
probability distribution induced by $\mu^m$ on $q'=(q_2, \dots, q_m)$.
\begin{Def} Let $J$ be a finite set with $\,|J| = m$.
Consider a function $\Phi:\, \db R^J\to\db R$ on $\R^J$  which we will identify with
$\R^m$. Function $\Phi$ is called $J$-monotonic if it satisfies the following conditions:
\psn
{\rm (1)} for any $r\in\db R^m_+$ and any $q\in \db R^m$,
\begin{equation}\label{Monoton1}
\Phi(q+r)\geq \Phi(q);
\end{equation}
\psn
{\rm(2)} moreover, for $e=e_1 + \dots + e_m\in \db R^m$, for any $q\in\db R^m$
and for any $t>0$
\begin{equation}\label{Monoton2}
\Phi(q + t \cdot e) - \Phi(q) \geq t.
\end{equation}
\end{Def}

  It is convenient to introduce the notion of $J$-monotonic operators considered
as  quadratic forms. In the following definition, we use the same notations as above.

\begin{Def} Let $\cH$ be a Hilbert space. A family of self-adjoint operators
$B:\cH \times \R^J \to \cH$ is called $J$-monotonic if, for any vector
$f\in\cH$ with $\|f\|=1$, the function $\Phi_f: \R^J \to \R$ defined by
$$
\Phi_f(q) = (B(q)f, f)
$$
is monotonic.
\end{Def}

  In other words, the quadratic form $Q_{B(q)}(f):= (B(q)f,f)$ as function of $q\in\R^J$
is non-decreasing in any $q_j$, $j=1, \dots, |J|$, and
$$
(B(q+t\cdot e)f, f) - (B(q)f,f) \geq t\cdot \|f\|^2.
$$
\pmn
\begin{Rem} By virtue of the variational principle for self-adjoint operators,
if an operator  family $H(q)$ in a finite-dimensional Hilbert space $\cH$ is
$J$-monotonic, then each eigen-value $E_k^{B(q)}$ of $B(q)$ is a $J$-monotonic function.
\end{Rem}

\begin{Rem} If  $H(q), \, q\in\R^J$,  is a $J$-monotonic  operator  family
in Hilbert space $\cH$, and $H_0:\cH\to\cH$ is an arbitrary self-adjoint operator, then
the family $H_0 + H(q)$ is also $J$-monotonic.
\end{Rem}

   This explains why the notion of monotonicity is relevant to spectral theory of random operators. Note
also, that this property can be easily extended to physically interesting examples where
$\cH$ has infinite dimension, but
$H(q)$ have, e.g., compact resolvent, as in the case of Schr\"{o}dinger operators in a finite cube with
Dirichlet b.c. and with bounded potential, so the respective spectrum is pure point, and
even discrete.

\begin{Thm}[Stollmann, \cite{St}]\label{Stollmann} Let $J$ be a finite index set,
$|J|=m$,
$\mu$ be a probability measure on $\db R$, and $\mu^m$ be the product measure on $\R^J$
with marginal measures $\mu$. If the function $\Phi:\, \db R^J \to \db R$ is
$J$-monotonic, then for any open interval $I\subset \db R$ we have
$$
\mu^m\myset{ q:\, \Phi(q) \in I } \leq m \cdot s(\mu, |I|).
$$
\end{Thm}
\proof
Let $I=(a,b)$, $b-a=\eps>0$, and consider the set
$$
A = \myset{ q:\, \Phi(q) \leq a  }.
$$
Furthermore, define recursively sets $A^\eps_j$, $j=0, \dots, m$, by setting
$$
A^\eps_0 = A, \; A^\eps_j = A^\eps_{j-1} + [0,\eps]e_j :=
\, \myset{ q+te_j:\, q\in A^\eps_{j-1}, \,\, t\in[0,\eps]  }.
$$
Obviously, the sequence of sets $A^\eps_j$, $j=1, 2, ...$, is increasing with $j$. The
monotonicity property implies
$$
\myset{ q:\, \Phi(q) < b  } \subset A^\eps_m.
$$
Indeed, if $\Phi(q) < b$, then for the vector $q':=q- \eps\cdot e$ we have by (2):
$$
\Phi(q') \leq \Phi(q' + \eps \cdot e) - \eps
= \Phi(q) - \eps \leq b - \eps \leq a,
$$
meaning that $q'\in \myset{ \Phi \leq a} = A$ and, therefore,
$$
 q = q' + \eps\cdot e \in A^\eps_m.
$$
Now, we conclude that
$$
\myset{ q:\, \Phi(q) \in I  } = \myset{ q:\, \Phi(q) \in (a,b)  }
$$
$$
= \myset{ q:\, \Phi(q) < b  } \setminus \myset{ q:\, \Phi(q) \leq a  }
\subset A^\eps_m \setminus A.
$$
Furthermore,
$$
\mu^m\myset{ q:\, \Phi(q) \in I } \leq
\mu^m\left( A^\eps_m \setminus A \right)
$$
$$
= \mu^m\left( \bigcup_{j=1}^m\left(  A^\eps_j \setminus A^\eps_{j-1} \right)
\right)
\leq \sum_{j=1}^m \mu^m\left( A^\eps_j \setminus A^\eps_{j-1} \right).
$$
For $q'\in\db R^{m-1}$, set
$$
I_1(q') = \myset{ q_1\in\db R:\, (q_1,q')\in A^\eps_1\setminus A}.
$$
By definition of set $A^\eps_1$, this is an interval of length not bigger than $\eps$.
Then we have
\begin{equation}\label{onestep}
\mu^m( A^\eps_1 \setminus A) = \int d\mu^{m-1}(q') \, \int_{I_1} d\mu(q_1)
\leq s(\mu, \eps).
\end{equation}
Similarly, we obtain for $j=2, \dots, m$
$$
\mu^m( A^\eps_j \setminus A^\eps_{j-1}) \leq s(\mu,\eps),
$$
yielding
$$
\mu^m\myset{ q:\, \Phi(q) \in I } \leq
\sum_{j=1}^m \mu^m( A^\eps_j \setminus A^\eps_{j-1})
\leq m \cdot c(\mu, \eps). \;\;\;\mathQED
$$

  Now, taking into account the above Remark 1, Stollmann's theorem yields immediately the
following estimate.

\begin{Thm} Let $H_\Lam$ be an LSO with random potential $V(x;\om)$
in a finite box $\Lam\subset\Z^d$ with Dirichlet b.c.,
and $\Sigma(H_\Lam)$ its spectrum, i.e. the collection of its eigen-values
$E^{(\Lam)}_j$, $j=1, \dots, |\Lam|$. Assume that r.v. $V(x;\cdot)$ are i.i.d. with
marginal distribution function $F_V$ satisfying
$$
s(\eps) = \sup_{a\in \R} \,\, (F_V(a+\eps) - F_V(a) )  < \infty.
$$
Then
$$
\pr{ \dist(\Sigma(H_\Lam(\om), E) \leq \eps } \leq  |\Lam|^2 s(\eps).
$$
\end{Thm}

\section{Extension to multi-particle systems}

   Results of this section have been obtained by the author and Y. Suhov \cite{ChS}.

Let $N> 1$ and $d\geq 1$ be two positive integers and consider a random LSO $H=H(\om)$
which can be used, in the framework of tight-binding approximation, the as the
Hamiltonian of a system of $N$ quantum particles in $\Z^d$ with random external potential
$V$ and interaction potential $U$. Specifically, let $x_1, \ldots, x_N\in\Z^d$ be
positions of quantum particles in the lattice $\Z^d$, and $\ux = (x_1, \ldots, x_N)$. Let
$\{V(x;\om), \, x\in\Z^d\}$ be a random field on $\Z^d$ describing the external potential
acting on all particles, and $U:\, (x_1,\ldots, x_N) \mapsto \R$ be the interaction
energy of the particles. In physics, $U$ is usually to be symmetric function of its $N$
arguments $x_1, \ldots, x_N\in\Z^d$. We will assume in this section that the system in
question obeys either Fermi or Bose quantum statistics, so it is convenient to assume
$U$ to be symmetric. Note, however, that the results of this section can be extended,
with natural modifications, to more general interactions $U$. Further, in \cite{ChS} $U$
is assumed to be finite-range interaction:
$$
\supp \, U \subset \{\ux:\, \max (|x_j-x_k|\leq r)\}, \; r<\infty.
$$
Such an assumption is required in the proof of Anderson localization for multi-particle
systems, however, it is irrelevant to the Wegner--Stollmann estimate we are going to
discuss below.

  Now, let $H$ be as follows:
$$
(H(\om) f)(\ux) = \sum_{j=1}^N \, \left( \Delta^{(j)} + V(x_j;\om)\right)
+ U(\ux),
$$
where $\Delta^{(j)}$ is the lattice Laplacian acting on the $j$-th particle, i.e.
$$
\Delta^{(j)} =  \truc{\one}{}{1} \otimes \ldots \otimes \truc{\Delta}{}{j} \otimes
\ldots \otimes \truc{\one}{}{N}
$$
acting in Hilbert space $ \ell^2(\Z^{Nd})$. For any finite "box"
$$
\Lam = \Lam^{(1)} \times \ldots \times \Lam^{(N)} \subset \Z^{Nd}
$$
one can consider the restriction, $H_\Lam(\om)$, of $H(\om)$ on $\Lam$ with Dirichlet
b.c. It is easy to see that the potential
$$
W(\ux) = \sum_{j=1}^N \, V(x_j;\om) + U(\ux)
$$
is no longer an i.i.d. random field on $\Z^{Nd}$, even if $V$ is i.i.d. Therefore,
neither Wegner's nor Stollmann's {\it estimate} does not apply {\it directly}. But, in
fact, Stollmann's lemma {\it does} apply to multi-particle systems, virtually in the same
way as to single-particle ones.
\begin{Thm} Assume that r.v. $V(x;\cdot)$ are i.i.d. with
marginal distribution function $F_V$ satisfying
$$
s(\eps) = \sup_{a\in \R} \,\, (F_V(a+\eps) - F_V(a) )  < \infty.
$$
Then
$$
\pr{ \dist(\Sigma(H_\Lam(\om), E) \leq \eps } \leq  |\Lam| \cdot M(\Lam)\cdot s(\eps),
$$
with
$$
M(\Lam) = \sum_{j=1}^N \, \card\, \Lam^{(j)} .
$$
\end{Thm}

\proof
Fix $\Lam$ and consider the union of all lattice points in $\Z^d$ which belong to the
single-particle projections $\Lam^{(j)}$, $j=1, \dots, N$:
$$
\cX(\Lam) = \bigcup_{j=1}^N \, \Lam^{(j)} \subset \Z^d.
$$
Now we can apply Stollmann's lemma to $H_\Lam$ by taking the index set  $J = \cX(\Lam)$
and auxiliary probability space $\R^J$. Indeed, the random potential
$\hat V(\ux;\om) := V(x_1;\om) + \dots + V(x_N;\om)$  can be re-written as follows:
$$
V(x_1;\om) + \dots + V(x_N;\om)
= \sum_{y\in \cX(\Lam)} c(\ux, y) V(y;\om)
$$
with integer coefficients $c(\ux,y)$ such that

\begin{equation}\label{multi}
c(\ux, y) \geq 0, \; \sum_{y\in \cX(\Lam)} c(\ux, y) = N.
\end{equation}
For example, if $N=2$, one can have either $V(x_1,\om) + V(x_2;\om)$ with $x_1\neq x_2$,
in which case we have
$$
c(\ux, y) =
\begin{cases}
  1, & \text{ if } y=x_1 \text{ or } y=x_2 \\
  0, & otherwise
\end{cases}
$$
or $V(x_1;\om) + V(x_1;\om)=2V(x_1;\om)$ for "diagonal" points $(x_1,x_2)$, where
$$
c(\ux, y) =
\begin{cases}
  2, & \text{ if } y=x_1  \\
  0, & otherwise
\end{cases}
$$
In any case, as shows (\ref{multi}), random potential at $\ux\in\Lam$ is a linear
function of one or more coordinates in the auxiliary space $\R^J$ growing at rate
$\geq Nt\geq t$ along the principal diagonal $\{q_1 = q_2 = \dots = q_{|J|} = t\in\R\}$. Hence,
the operators of multiplication by $\hat V(\ux;\om)$ form a $J$-monotonic family, and, by
virtue of Remark 2, the same holds for $H = H_0 + U + \hat V(\om)$, just as in the
single-particle case (and even "better", for
$N>1$ !). By Theorem 2, this implies immediately the estimate
$$
\pr{ \dist(\Sigma(H_\Lam(\om), E) \leq \eps } \leq  |\Lam|^2 s(\mu,\eps). \qquad \mathQED
$$

   It is not difficult to see that the same argument, with obvious notational
modifications, applies to Fermi and Bose lattice quantum systems, i.e. to restrictions of
$H$ to the subspaces of symmetric (Bose case) or anti-symmetric (Fermi case) functions
of $N$ arguments $x_1, \dots, x_N$ on $(\Z^d)^{N}$.

\section{Extension to correlated random variables}

   Now let $\mu^m$ be  a measure on $\db R^m$ with marginal distributions of order $m-1$,
$$
\mu^{m-1}_j(q'_{\neq j}) = \mu^{m-1}_j(q_1, \dots, q_{j-1}, q_{j+1}, \dots, q_m),
\; j=1, \dots, m,
$$
and conditional distributions $\mu^{1}_j(q_j \cnd q'_{\neq j})$ on $q_j$ given all
$q_k, k\neq j$.
For every $\eps>0$, define the following quantity:
$$
C_1(\mu^m, \eps) = \max_{j} \,\sup_{a\in\db R} \,\,
\int d\mu^{m-1}(q'_{\neq j}) \, \int_{a}^{a+\eps} d\mu(q_1|q'_{\neq j})
$$
and assume that $C_1(\mu,\eps)$ is finite:
\begin{equation}\label{CondA}
 \max_{j} \,\sup_{a\in\db R} \,\,
\int d\mu^{m-1}(q'_{\neq j}) \, \int_{a}^{a+\eps} d\mu(q_1|q'_{\neq j}) < \infty.
\end{equation}
\begin{Rem}  As a simple sufficient condition of finiteness
of  $C_1(\mu,\eps)$, one can use, e.g., a uniform continuity (but not necessarily {\bf
absolute continuity} !) of the single-point conditional distributions,
$$
\max_{j}\,\sup_{ q'_{\neq j} } \, \sup_{a\in\db R} \,\,
 \int_{a}^{a+\eps} d\mu(q_j|q'_{\neq j}) \leq C_2(\mu^m,\eps) < \infty
$$
or even the existence and uniform boundedness of the {\bf density } $p(q_j|q'_{\neq j})$
of these conditional distributions:
$$
\sup_{q_j\in\db R} p(q_j|q'_{\neq j}) \leq C_3(\mu^m,\eps).
$$
\end{Rem}
\pn
\begin{Rem}  In applications to localization problems, the aforementioned
continuity moduli $C_1(\mu^m,\eps)$, $ C_2(\mu^m,\eps)$,  $C_3(\mu^m,\eps)$ need to decay
not too slowly as $\eps\to 0$. A power decay of order $O(\eps^\beta)$ with $\beta>0$ is
certainly sufficient, but it can be essentially relaxed. For example, it suffices to have
an upper bound of the form
$$
C_1\left(\mu^m, e^{-L^{\beta}}\right) \leq Const \cdot  L^{-B},
$$
uniformly for all sufficiently large $L>0$ with some (arbitrarily small) $\beta>0$ and
with $B>0$ which should sufficiently big, depending on the specific spectral problem.
\end{Rem}

\pn
 Using notations of the previous
section, one can formulate the following generalization of Stollmann's lemma.
\begin{Lem}\label{ExtStollmann}
Let $\Phi:\, \db R^J \to \db R$, $\R^J \cong \R^m$, be a $J$-monotonic function and
$\mu^m$ a probability measure on $\R^m\cong \R^J$ with $C_1(\mu^m,\eps)<\infty$. Then for any
interval
$I\subset \db R$ of length $|I|=\eps>0$, we have
$$
\mu^m\myset{ q:\, \Phi(q) \in I } \leq m \cdot C_1(\mu, \eps).
$$
\end{Lem}
\proof
We proceed as in the proof of Stollmann's lemma and introduce in $\db R^m$ the sets
$A = \myset{ q:\, \Phi(q) \leq a  }$ and $A^\eps_j$, $j=0, \dots, m$. Here, again,
we have
$$
\myset{ q:\, \Phi(q) \in I  } = \subset A^\eps_m \setminus A
$$
and
$$
\mu^m\myset{ q:\, \Phi(q) \in I }
\leq \sum_{j=1}^m \mu^m\left( A^\eps_j \setminus A^\eps_{j-1} \right).
$$
For $q'_{\neq 1}\in\db R^{m-1}$, we set
$$
I_1(q'_{\neq 1}) = \myset{ q_1\in\db R:\, (q_1,q'_{\neq 1})\in A^\eps_1\setminus A}.
$$
Furthermore, we come to the following upper bound which generalizes (\ref{onestep}):
\begin{equation}\label{onestepgen}
\mu^m( A^\eps_1 \setminus A) = \int d\mu^{m-1}(q') \, \int_{I_1} d\mu(q_1|q')
\leq C_1(\mu, \eps).
\end{equation}
Similarly, we obtain for $j=2, \dots, m$
$$
\mu^m( A^\eps_j \setminus A^\eps_{j-1}) \leq C_1(\mu,\eps),
$$
yielding
$$
\mu^m\myset{ q:\, \Phi(q) \in I } \leq
\sum_{j=1}^m \mu^m( A^\eps_j \setminus A^\eps_{j-1})
\leq m \cdot C_1(\mu, \eps). \;\;\;\mathQED
$$

\section{Application to Gaussian random fields}

  Let $V(x,\omega), x\in\Z^d$, $d\geq 1$, be a regular stationary Gaussian field of
zero mean on the lattice $\Z^d$. The regularity implies that the field $V(\cdot, \omega)$
is non-deterministic, i.e. the conditional probability distribution of $V(0,\cdot)$ given
$\{V(y), y\neq 0\}$ is Gaussian with strictly positive variance. In other terms, the r.v.
$V(0,\cdot)$, considered as a vector in the Hilbert space $\cH_{V,\Z^d}$ generated by
linear combinations of all $V(x,\cdot)$, $x\in\Z^d$, with the scalar product
$$
(\xi,\eta) = \esm{\xi\, \eta},
$$
does not belong to the subspace $\cH_{V,\Z^d \setminus \{0\}}$:
$$
\| V(0,\cdot) - \Pi_{\cH_{V,\Z^d  \setminus \{0\}}}  V(0,\cdot) \|^2
= \tilde \sigma_0^2 >0,
$$
where
$$
\Pi_{\cH_{V,\Z^d}} \xi
= \esm{\xi \, \Big|\, V(x,\cdot), x\in\Z^d \setminus \{0\}}.
$$
Furthermore, for any subset $\Lam \subseteq \Z^d \setminus \{0\}$,
$$
\| V(0,\cdot) - \Pi_{\cH_{V, \Lam}}  V(0,\cdot) \|^2 \geq \tilde \sigma_0^2,
$$
since
$$
\cH_{V,\Lam} \subset \cH_{V,\Z^d \setminus \{0\}}.
$$
Therefore, the conditional variance of $V(0,\cdot)$ given any non-zero number of values
of $V$ outside $x=0$ is bounded from below by $\tilde \sigma_0^2$. Respectively, the
conditional probability density of $V(0,\cdot)$, for any such nontrivial condition is
uniformly bounded by $(2\pi\tilde \sigma_0^2)^{-1/2} <\infty$. Now a direct application
of Lemma \ref{ExtStollmann} leads to the following statement.
\begin{Thm}
   Let $\Lam \subset \Z^d$ be a finite subset of the lattice, and
$\Lam' \subset \Z^d \setminus \Lam$ any subset  disjoint with
$\Lam$ ($ \Lam'$  may be empty). Consider a family of  LSO
$H_{\Lam}(\om)$ with Gaussian random potential
$V(\om)$ in $\Lam$, with Dirichlet b.c. on $\pt \Lam$. Then for any interval
$I\subset \db R$ of length $\eps>0$, we have
$$
\pr{ \Sigma( H_{\Lam}) \cap I \neq \emptyset\,|\, V(y,\cdot), y \in \Lam'}
\leq C(V) \, |\Lam|^2 \, \eps,
$$
where the constant $C(V)<\infty$ whenever the Gaussian field $V$ is non-deterministic.
\end{Thm}

\section{Application to Gibbs fields with continuous spin}

   Apart from Gaussian fields, there exist several classes of random lattice fields for
which the hypothesis of Lemma \ref{ExtStollmann} can be easily verified. For example,
conditional distributions of Gibbs fields are given explicitly in terms of their
respective interaction potentials. Specifically, consider a lattice Gibbs field
$s(x,\omega)$ with bounded continuous spin,
$$
s: \Omega \times \Z^d \to \cS = [a,b] \subset \db R
$$
generated by a short-range, bounded, two-body interaction potential $u(\cdot,\cdot)$. The
spin space is assumed to be equipped with the Lebesgue measure $ds$. In other words,
consider the formal Hamiltonian
$$
H(s) = \sum_{x\in \Z^d} h(x)
+ \sum_{x\in \Z^d} \sum_{|y-x|\leq R} u_{|x-y|}(s(x), u(y)),
$$
where $h:\cS\to \db R$ is the self-energy of a given spin. The interaction potentials
$u_{|x-y|}(s(x),s(y))$ vanish for $|x-y|>R$ and are uniformly bounded:
$$
\max_{l\leq R}\sup_{s,t\in\cS} |u_{l}(s,t)| <\infty.
$$
Then for any lattice point $x$ and any configuration $s' = s'_{\neq x}$ of spins outside
$\{x\}$, the {\it single-site} conditional distribution of $s(x)$ given the external configuration
$s'$ admits a {\it bounded density}
$$
p(s_x \,|\,s'_{\neq x})
= \frac{ e^{-\beta U(s_x | s'})}{ \Xi(\beta, s') }
= \frac{ e^{-\beta U(s_x | s'})}{ \int_{\cS} e^{-\beta U(t | s'}) \, dt}
$$
with
$$
U(s_x | s') := \sum_{y:\, |y-x|\leq R} u_{|x-y|}(s_x,s'_y)
$$
satisfying the upper bound
$$
|U(s_x | s')| \leq (2R+1)^d  \sup_{s,t\in \cS} |u_{l}(s,t)| <\infty.
$$
A similar property is valid for sufficiently rapidly decaying long-range interaction
potentials, for example, under the condition
\begin{equation}\label{GibbsSummable}
 \sup_{s,t\in\cS} \, |u_{|y|}(s,t)| \leq \frac{Const }{ |y|^{d+1+\delta}}, \; \delta>0.
\end{equation}
as well as for more general, but still uniformly summable many-body interactions. Here is
one possible Wegner--Stollmann-type result concerning such random potentials.
\begin{Thm}
   Let $\Lam \subset \Z^d$ be a finite subset of the lattice,
$\Lam' \subset \Z^d \setminus \Lam$ any subset  disjoint with $\Lam$ ($ \Lam'$
may be empty), and let $s(x,\omega)$ be a Gibbs field in $\Lam$ with continuous spins
$s\in\cS=[a,b]$ generated by a two-body interaction potential $u_l(s,t)$ satisfying condition
(\ref{GibbsSummable}), with any b.c. on $\Z^d \setminus \Lam$. Consider a LSO $H_{\Lam}$
with random potential $V(x,\omega)= s(x,\omega)$. Then for any interval $I\subset \db R$
of length $\eps>0$, we have
$$
\pr{ \Sigma( H_{\Lam}) \cap I \neq \emptyset\,|\, V(y,\cdot), y \in \Lam'}
\leq C(V) \, |\Lam|^2 \, \eps,\; C(V)<\infty.
$$
\end{Thm}

   In the case of unbounded spins and/or interaction potentials, the uniform
boundedness of conditional single-spin distributions does not necessarily hold,  since
the energy of interaction of a given spin $s(0)$ with the external configuration $s'$ may
be arbitrarily large (depending on a particular form of interaction) and even {\it
infinite}, if $s'(y) \to \infty$ too fast. In such situations, our general condition
(\ref{CondA}) may still apply, provided that rapidly growing configurations $s'$ have
sufficiently small probability, so that the outer integral in the r.h.s. of (\ref{CondA})
converges.

\section{Conclusion}

  Wegner--Stollmann-type estimate of the density of states in finite volumes is a key ingredient of
the MSA of spectra of random Schr\"{o}dinger (and some other) operators. The proposed simple
extension of  Stollmann's lemma shows that a very general assumption on correlated random
fields generating potential rules out an abnormal  accumulation of eigen-values in finite
volumes. This extension applies also to multi-particle systems \cite{ChS2}.

\end{document}